\documentclass[]{aastex63}
\usepackage{graphicx}

\renewcommand\email\texttt

\usepackage{amsmath}
\usepackage{natbib}
\usepackage{CJKutf8}
\submitjournal{ApJ}

\shorttitle{structural and kinematical properties of NGC 5634}
\shortauthors{Wang et al.}
\graphicspath{{./}{figures/}}
\begin{document}

\title{The structural and kinematical properties of NGC 5634, a globular cluster associated with the Sagittarius dwarf galaxy? }

\correspondingauthor{Jundan Nie}
\email{jdnie@nao.cas.cn}

\author[0009-0006-8509-8888]{Shouzhi Wang\ 
\begin{CJK}{UTF8}{gbsn}(王守智)\end{CJK}}
\affil{Institute for Frontiers in Astronomy and Astrophysics, Beijing Normal University, Beijing 102206, P.R. China}
\affil{School of Physics and Astronomy, Beijing Normal University, Beijing 100875, P.R. China}
\affil{Key Laboratory of Space Astronomy and Technology, 
National Astronomical Observatories, 
Chinese Academy of Sciences, 
Beijing 100101, PR China}
%\nocollaboration{2}

\author[0000-0001-6590-8122]{Jundan Nie\ 
\begin{CJK}{UTF8}{gbsn}(聂俊丹)\end{CJK}}
\affil{Key Laboratory of Space Astronomy and Technology, 
National Astronomical Observatories, 
Chinese Academy of Sciences, 
Beijing 100101, PR China}
%\nocollaboration{2}

\author[0000-0003-3168-2617]{Biwei Jiang\ 
\begin{CJK}{UTF8}{gbsn}(姜碧沩)\end{CJK}}
\affil{Institute for Frontiers in Astronomy and Astrophysics, Beijing Normal University, Beijing 102206, P.R. China}
\affil{School of Physics and Astronomy, Beijing Normal University, Beijing 100875, P.R. China}

\author[0000-0003-3347-7596]{Hao Tian\ 
\begin{CJK}{UTF8}{gbsn}(田浩)\end{CJK}}
\affil{Key Laboratory of Space Astronomy and Technology, 
National Astronomical Observatories, 
Chinese Academy of Sciences, 
Beijing 100101, PR China}

\author[0000-0002-1802-6917]{Chao Liu\ 
\begin{CJK}{UTF8}{gbsn}(刘超)\end{CJK}}
\affil{Key Laboratory of Space Astronomy and Technology, 
National Astronomical Observatories, 
Chinese Academy of Sciences, 
Beijing 100101, PR China}

\author[0009-0009-9527-0444]{Ying-Hua Zhang\ 
\begin{CJK}{UTF8}{gbsn}(张颖华)\end{CJK}}
\affil{Key Laboratory of Space Astronomy and Technology, 
National Astronomical Observatories, 
Chinese Academy of Sciences, 
Beijing 100101, PR China}
\affil{School of Astronomy and Space Science, University of Chinese Academy of Sciences, Beijing 
100048, P.R China}

\begin{abstract}

We investigate the origin of NGC 5634 through a comprehensive analysis of its morphology, kinematics and dynamics. Utilizing data from the DESI Legacy Survey, we refined its fundamental parameters (age $\tau = 12.8 \pm 0.3$ Gyr, metallicity $[Fe/H] = -1.8 \pm 0.1$ dex, distance modulus $dm = 17.0 \pm 0.1$ mag) and constructed matched-filter template based on the combination of these parameters to search for extra-tidal structures. However, no significant features were detected above a $3\sigma$ signal-to-noise threshold, which limits our ability to further investigate the association between NGC 5634 and the Sagittarius (Sgr) stream based on morphological evidence. Incorporating GAIA data, we further examine the orbital path of NGC 5634. We found that its orbit only briefly intersects with the Sgr stream and diverges significantly over long-term integrations. This behavior contrasts with that of confirmed Sgr-associated clusters, whose orbits remain closely aligned with the stream throughout their orbital evolution. Additionally, NGC 5634 exhibits a relatively shorter semi-major axis and smaller apocenter and pericenter distances compared to Sgr clusters. These orbital characteristics are more consistent with clusters associated with the Gaia-Sausage-Enceladus (GSE) or the Helmi streams. From a dynamical perspective, in the $L_z$–$E$ space, NGC 5634 is also distinctly different from Sgr clusters and aligns more closely with the GSE and Helmi regions. Taken together, these findings do not support a strong connection between NGC 5634 and the Sgr dSph, but instead suggest a potential association with another progenitor system, such as GSE or Helmi stream. Nevertheless, further evidence is needed to definitively establish its origin.

\end{abstract}


\section{Introduction}\label{introduction}

Globular clusters (GCs) are ancient stellar systems that hold critical information about the formation and evolution of their host galaxies. In the context of our Galaxy, GCs serve as key tracers of both the in situ formation of the Galactic halo and the accretion of external systems. Over the last few decades, compelling evidence has demonstrated that a substantial fraction of the galactic GC population originated in satellite galaxies that were later accreted and disrupted by the galactic tidal forces. Multiple satellite accretion events have been identified in the galactic history, such as Kraken, Sequoia, Helmi, Gaia-Sausage-Enceladus (GSE), and the Sagittarius dwarf spheroidal galaxy (Sgr dSph). Among these, the Sgr dSph and its associated stellar stream (Sgr stream) represent one of the most significant accretion events in the history of the Galactic halo because its merging is still ongoing.

The Sgr stream spans a vast region surrounding the Galactic halo, with multiple wraps extending over scales of several tens to approximately 100 kpc, providing compelling evidence of a significant accretion event \citep[see, e.g.,][and references therein]{1994Natur.370..194I,2001ApJ...547L.133I,2002ApJ...569..245N,2003ApJ...599.1082M,2006ApJ...642L.137B,2007ApJ...668..221N,2010ApJ...712..516N,2010ApJ...721..329C,2014MNRAS.437..116B}. These stars and clusters in Sgr stream retain kinematic and chemical signatures of their origin, enabling the identification of accreted components \citep{2010ApJ...714..229L,2021MNRAS.501.2279V}. Several GCs, such as NGC 6715 (M54), Ter 7, Ter 8, Arp 2, Pal 12 and Whiting 1, have been definitively linked to the Sgr dSph based on their spatial, kinematical, and chemical properties \citep{2010ApJ...718.1128L,2018ApJ...862...52S,2019MNRAS.484.2832V,2019A&A...630L...4M}. However, the association of other GCs, including NGC 5634, remains an open question.

NGC 5634 is a relatively metal-poor ([Fe/H] = -1.876) and dynamically old GC that has been suggested to originate from the Sgr dSph due to its spatial proximity to the Sgr stream and its position and radial velocity being compatible with the hypothesis that the cluster retains the motion characteristics of the Sgr dSph \citep{2002AJ....124..915B,2010ApJ...718.1128L}. \cite{2002AJ....124..915B} presented the first detailed Color–Magnitude Diagram (CMD) analysis of NGC 5634, which highlighted its steep Red Giant Branch and extended blue Horizontal Branch. These branches show a strong resemblance to those of the Sgr cluster Ter 8. Additionally, the prominent blue Horizontal Branch population closely matches the stellar population observed in the Sgr stream \citep{2002ApJ...569..245N}. Further spectroscopic studies \citep[e.g.,][]{2015A&A...579A.104S,2017A&A...600A.118C} have revealed abundance patterns in NGC 5634 that include the Na-O anticorrelation typical of GCs but also exhibit similarities to the chemical enrichment patterns observed in Sgr stars. Thus, building on the above-mentioned studies, \cite{2020A&A...636A.107B} identified NGC 5634 as a strong candidate for being an accreted GC associated with the Sgr dSph. Moreover, they thought the cluster has a position and proper motion that are consistent with the ancient arm located at a distance of 20$-$25 kpc, as previously identified by \cite{2010ApJ...721..329C}.

Determining the origin of a GC like NGC 5634 requires a combination of methods. Traditional approaches typically involve comparing its spatial distribution, proper motions, radial velocities, and age-metallicity relationship to assess whether it belongs to a known progenitor system. These methods have proven effective in many cases, particularly with the availability of high-precision astrometric data from Gaia \citep{2016A&A...595A...1G,2021A&A...649A...1G}. However, they are not without limitations. GCs may have undergone significant dynamical perturbations, resulting in deviations from their original positions, while proper motions and radial velocities may exhibit considerable dispersion, potentially overlapping with those of other sources. Furthermore, in the metal-poor, old regime, clusters originating from various systems tend to cluster together, making it difficult to distinguish their origins effectively. Additionally, the chemical tagging method, which traces origins using abundance patterns, also faces challenges due to the overlap of chemical signatures among different accreted systems, particularly for elements with weak evolutionary trends like [$\alpha$/Fe] \citep{2019MNRAS.488.1235M}. Given these challenges, utilizing a combination of morphological, kinematical, and dynamical characteristics of GCs has become a crucial approach for collectively determining their origins.

In this work, we used deep photometric data from the DESI survey to investigate the potential extra-tidal structures of NGC 5634 with a matched filter method. This enabled us to analyze its structural orientation and assess its origin. Previous studies on the extra-tidal structures of Whiting 1 \citep{2022ApJ...930...23N} and NGC 4147 \citep{2024AJ....168..237Z} provided a solid foundation for our approach. We further examined the orbital characteristics of NGC 5634 and related clusters, focusing on their orbits relative to Sgr stream. Orbital parameters, including semi-major axis, eccentricity, and inclination, were analyzed to provide deeper insights into their kinematical properties. Additionally, we assessed the angular momentum and total energy distributions to determine whether NGC 5634 aligns with a specific stellar stream. These analyses collectively aim to enhance our understanding of NGC 5634 and the broader population of galactic GCs. In Section 2, we detail the determination of NGC 5634's fundamental, structural, and 6D parameters in phase space, as well as the methodology and implementation of the matched filter technique. Section 3 presents the results of the tidal structure analysis, examines the cluster’s orbit path and orbital parameters, discusses potential factors that may affect the orbit, and subsequently evaluates its distribution in energy–angular momentum space. Finally, Section 4 provides a comprehensive summary of our findings.

\section{Data and Method}\label{data_method}

\subsection{DESI Legacy Surveys data}
\label{DESIdata}

The DESI Legacy Imaging Surveys \citep{2019AJ....157..168D} comprise a combination of three public surveys: the Dark Energy Camera Legacy Survey (DECaLS), the Beijing–Arizona Sky Survey (BASS), and the Mayall $z$-band Legacy Survey (MzLS). Together, these surveys will observe approximately 14,000 square degrees of the sky with $|b| > 20^\circ$, using telescopes located at the Kitt Peak National Observatory and the Cerro Tololo Inter-American Observatory. The observations span three optical bands ($g$, $r$, and $z$), providing a comprehensive dataset for studying the cosmos at multiple wavelengths. The average 5$\sigma$ point-source depths for these surveys are $g$ = 24.7, $r$ = 23.9, and $z$ = 23.0 mag, respectively. However, these depths can vary depending on observing conditions, such as atmospheric seeing and sky brightness.

Despite the high-quality data provided by the DESI surveys, we found certain limitations in the DR9 and DR10 datasets. Specifically, for dense GCs, some regions were not photometered due to the use of Gaia star catalog positions for forced photometry, resulting in incomplete photometric measurements. This incompleteness led to data loss in CMD, particularly for faint main-sequence stars. To overcome this limitation, we opted to use the DR8 dataset for our analysis, as it provided more reliable and complete photometric data for the purposes of this study.

In DR8, six morphological types are defined: point sources (PSF), round exponential galaxies with a variable radius (REX), de Vaucouleurs (DEV) profiles (elliptical galaxies), exponential (EXP) profiles (spiral galaxies), composite profiles (COMP) combining de Vaucouleurs and exponential models, and the DUP type, which is set for Gaia sources that coincide with and have been fit by an extended source. Since our study focuses on stellar sources, we exclusively extracted the sources of PSF type from the dataset. To further ensure data quality, we applied additional criteria, selecting stars with PSF magnitudes in the range of 14 $<$ $r$ $<$ 24 and photometric errors of ${\sigma}_r, {\sigma}_g$ $<$ 0.2 mag. For the spatial field of analysis, we defined a $5^\circ \times 5^\circ$ region centered on NGC 5634 (see Table \ref{table1}). To correct for Galactic extinction, we applied reddening corrections to all magnitudes using $E(B-V)$ values from \cite{1998ApJ...500..525S}, with extinction coefficients of 3.214, 2.165, and 1.211 for the $g$, $r$, and $z$ bands, respectively \citep{2019AJ....157..168D}.

To further ensure the reliability of the dataset and minimize the impact of observational conditions, we conducted a completeness test for NGC 5634 and its surrounding regions. First, we retrieved optical images of NGC 5634 from the DESI imaging dataset. These images were processed using SExtractor \citep{1996A&AS..117..393B} for photometric measurements, with configuration parameters tuned based on the telescope's characteristics and iteratively adjusted to optimize the photometric results. This approach ensures that our output star catalog is broadly consistent with the official DESI catalog. Once reliable photometric parameters were established, we injected artificial stars with random positions into the images to evaluate completeness. These stars, with magnitudes ranging from 15 to 24 mag, were simulated using the point spread function (PSF) of NGC 5634. Approximately 1000 artificial stars were added per $1^\circ \times 1^\circ$ area. Photometry was then re-performed on the modified images, and the resulting catalogs were cross-matched with the input artificial star catalog to calculate detection efficiency in each magnitude bin. This process was repeated 10 times to obtain the average detection efficiency for each magnitude, enabling an accurate assessment of the dataset’s overall completeness.

\begin{figure*}
\centering
\includegraphics[width=0.49\textwidth]{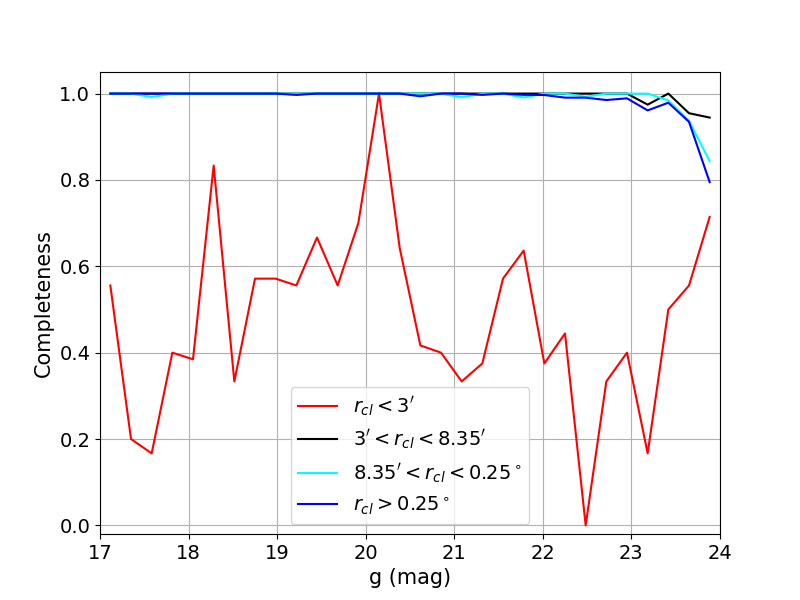}
\includegraphics[width=0.49\textwidth]{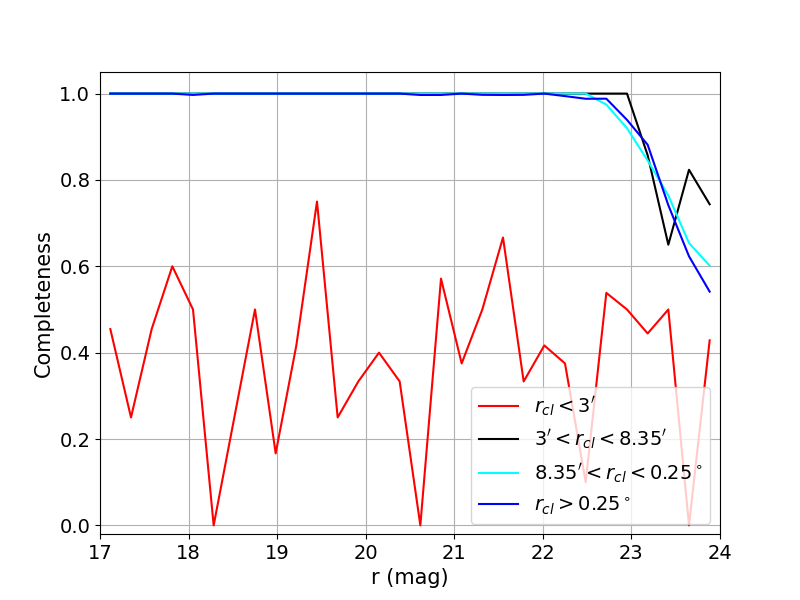}

\caption{Result of completeness test for our working field. Different colors represent the range of distances from the center of NGC 5634. The limit magnitudes in the $g$-band and $r$-band are 23.0 and 22.5 mag, respectively. }

\label{fig_completeness}
\end{figure*}

An observed effect was that the high stellar density in the central regions of NGC 5634 severely impacts photometric completeness. To quantify this effect, we divided the field into four distinct regions for completeness analysis: (1) the dense central core ($r_{cl}$ $<$ $3\arcmin$), (2) the intermediate region between the core and the tidal radius ($3\arcmin$ $<$ $r_{cl}$ $<$ $8.35\arcmin$), (3) the region beyond the tidal radius potentially containing extra-tidal structures ($8.35\arcmin$ $<$ $r_{cl}$ $<$ 0.25$^\circ$), and (4) the broad background field ($r_{cl}$ $>$ 0.25$^\circ$). Figure \ref{fig_completeness} illustrates the completeness in the $g$ and $r$ bands for each of these regions. The results clearly demonstrate that photometric completeness is significantly compromised in the dense central core (r$_{cl} < 3\arcmin$) due to crowding effects, resulting in a lower detection efficiency. Moreover, because the very central regions of GCs are so densely packed with stars, the stellar profiles blend together, making source detection extremely challenging. As a result, there is no discernible trend in the detection rates at the bright and faint ends; instead, they appear random (as demonstrated by the red lines in the Figure \ref{fig_completeness}). Hence, to mitigate the effects of incompleteness, we limited our analysis to regions beyond the dense central core ($r_{cl}$ $>$ $3\arcmin$). Furthermore, we restricted our study to magnitude ranges where completeness remained at 100 \%, ensuring the reliability of our results. These thresholds were determined to be $g$ = 23 and $r$ = 22.5 mag, respectively, beyond which detection efficiency declined rapidly.

\begin{deluxetable*}{cccccccccccc}
\tablenum{1}
\tablecaption{Filtered DESI photometric catalog within $5^\circ \times 5^\circ$ centered on NGC 5634. Col. 1-2: Equatorial coordinates in J2000.
Col. 3-5: Magnitude, photometric error, and extinction correction of g-band, respectively.
Col. 6-8: Same as Col. 3-5, but for r-band.
Col. 9-10: Proper motion in the RA direction and its error, in mas/yr.
Col. 11-12: Proper motion in the Dec direction and its error, in mas/yr. \label{table1}}
\tablewidth{0pt}
\tablehead{
\colhead{RA} & \colhead{Dec} & \colhead{gmag} & \colhead{gerr} & \colhead{gext} &
\colhead{rmag} & \colhead{rerr} & \colhead{rext}& \colhead{pmRA}& \colhead{pmRAerr} & \colhead{pmDec}& \colhead{pmDecerr}\\
\colhead{(J2000)} & \colhead{(J2000)} & \colhead{(mag)} & \colhead{(mag)} & \colhead{(mag)} &
\colhead{(mag)} & \colhead{(mag)} & \colhead{(mag)}& \colhead{mas/yr}& \colhead{mas/yr}& \colhead{mas/yr}& \colhead{mas/yr}
}
\startdata
214.93572	&-4.374075	&18.5255	&0.002	&0.1567	&17.4938	&0.0016	&0.1055	&-0.88413	&13.725545	&-0.529829	&25.178642\\
215.001782	&-4.374433	&19.9448	&0.0048	&0.1571	&18.8674	&0.0036	&0.1058	&-11.249587	&2.587103	&2.833001	&4.131823\\
214.991151	&-4.370406	&17.9877	&0.0013	&0.1572	&17.361	&0.0013	&0.1059	&-1.068638	&16.505714	&-9.043289	&21.377596\\
214.913443	&-4.367841	&17.8173	&0.0011	&0.1567	&17.0289	&0.0008	&0.1056	&-8.391708&	25.788933	&-1.15044	&42.932213\\
214.937845	&-4.368362	&20.7406	&0.0071	&0.157	&19.2357	&0.0036	&0.1058	&-6.668041	&2.189114	&-5.437576	&3.713203\\
......\\
\enddata
\tablecomments{Table 1 is published in its entirety in the machine-readable format. A portion is shown here for guidance regarding its form and content.
}
\end{deluxetable*}

%------------------------------------------------
%------------------------------------------------

%------------------------------------------------
\begin{figure}
\centering
\includegraphics[width=0.65\textwidth]{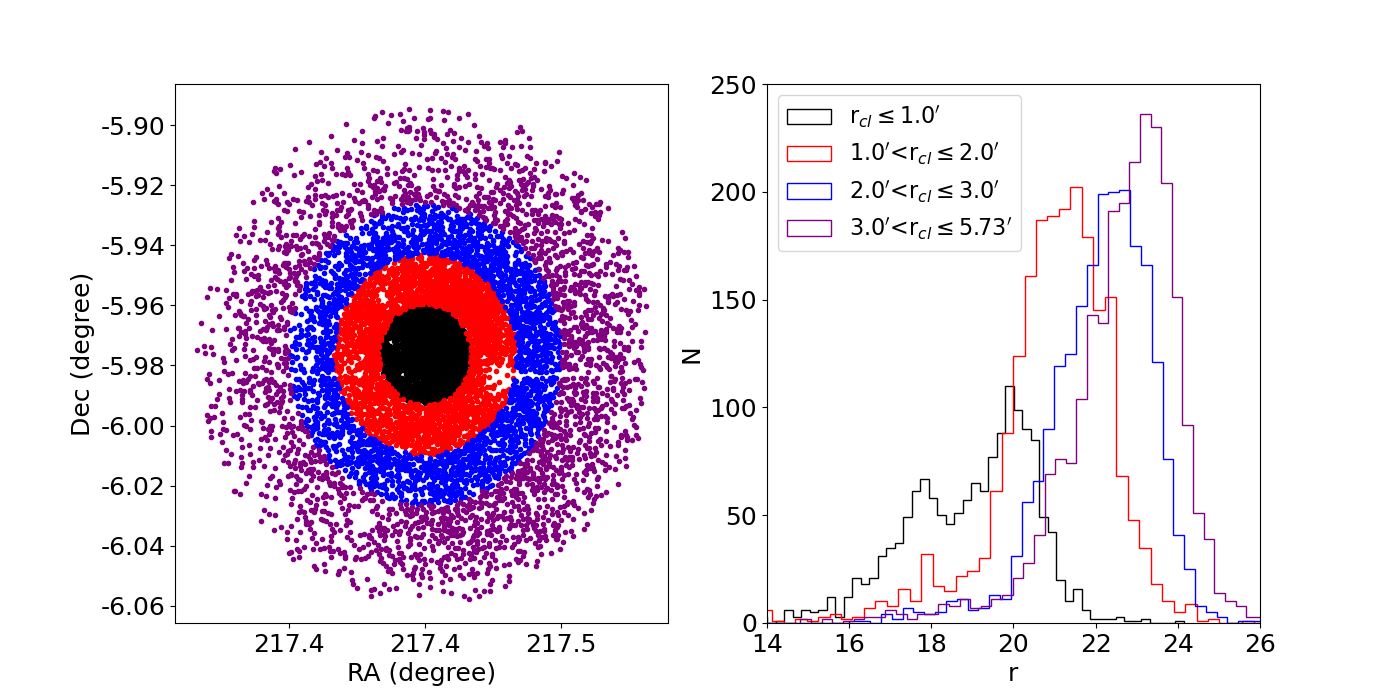}
\includegraphics[width=1.0\textwidth]{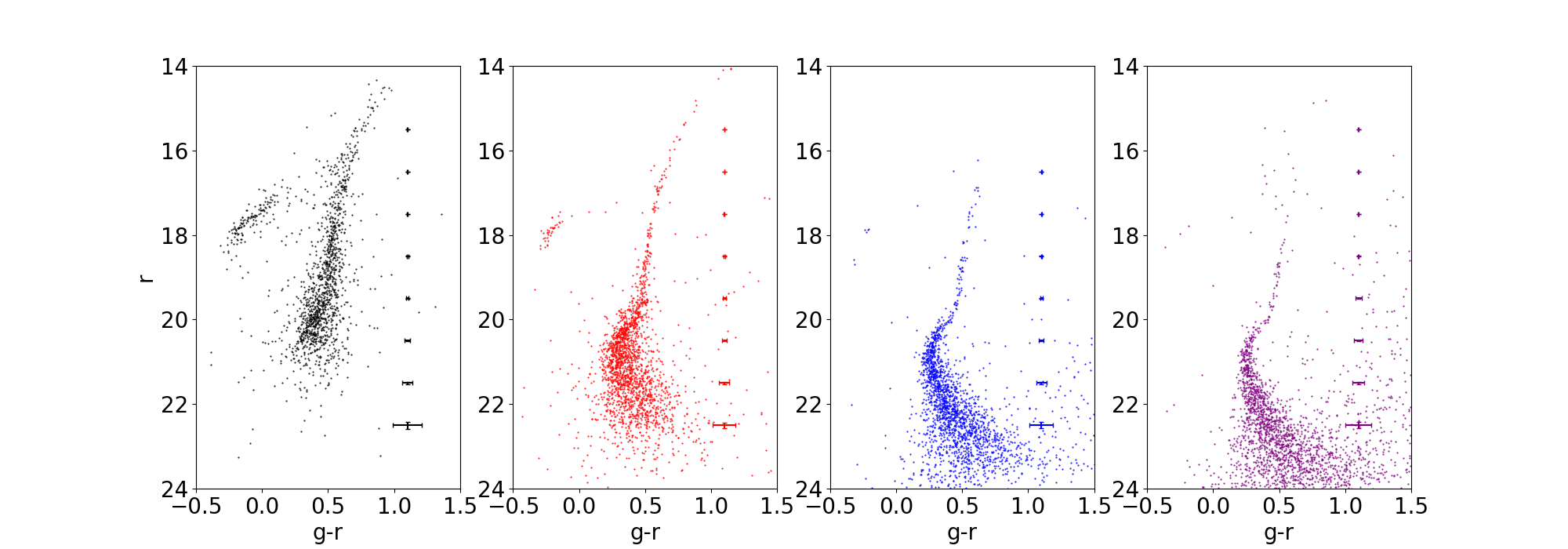}

\caption{Stellar distribution and corresponding CMDs for different regions within the tidal radius. The upper-left panel shows the multiple spatial distribution of stars centered on NGC 5634. The upper-right panel illustrates the magnitude distribution ($r$-band) for stars within each radius interval. The bottom panels present the CMDs for stars in each radius interval. Typical observational error bars are displayed at the right side.}

\label{fig_depth}
\end{figure}
%------------------------------------------------

%------------------------------------------------
\begin{figure}
    \centering
\includegraphics[width=0.95\textwidth]{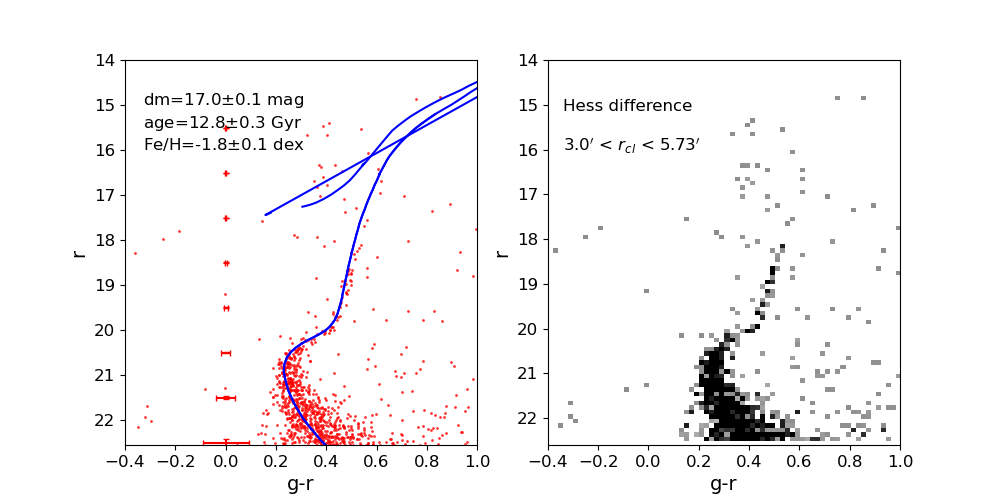}

\caption{Left panel: The observational CMD of NGC 5634. Red dots represent extinction-corrected stars within the radius of $3\arcmin$ $<$ $r_{cl}$ $<$ $5.73\arcmin$. Typical observational error bars are displayed at eight magnitude levels. The blue line shows the best-fit CMD model, illustrating an agreement with the observed data. Right panel: The background-subtracted Hess diagram for NGC 5634, serving as the final CMD template.}

\label{fig_cmd1}
\end{figure}
%------------------------------------------------

\subsection{Determination of the multiple parameters of NGC 5634 }
\label{allpara}
 
\subsubsection{Fundamental parameters}
\label{funpara}

The fundamental parameters of NGC 5634, including its age ($\tau$), metallicity ($Z$), and distance modulus ($dm$), were determined based on our analysis of DESI DR8 photometric data. These parameters were derived by fitting theoretical isochrones to the observed CMD. For this study, the CMD was constructed using the $g - r$ vs $r$ combination, as this effectively captures the full range of stellar evolutionary stages within the cluster. To correct for interstellar extinction, we applied the reddening values $E(B-V)$ provided by \cite{1998ApJ...500..525S} to the observed CMD.

The theoretical isochrones were generated using the CMD 3.8 input form, a web-based tool for stellar isochrones and their derivatives \footnote{stev.oapd.inaf.it/cgi-bin/cmd}. Specifically, we employed evolutionary tracks from the PARSEC version 1.2S+COLIBRI S$_{35}$ models \citep{2012MNRAS.427..127B, 2014MNRAS.445.4287T, 2014MNRAS.444.2525C, 2015MNRAS.452.1068C, 2019MNRAS.485.5666P}. The photometric system was set to DECam (AB magnitudes), with no dust effects applied to the stars and no extinction effects from interstellar dust included in the models, ensuring the isochrones were intrinsic and directly comparable to the extinction-corrected CMD. The initial mass function (IMF) was assumed to follow the power-law model described by \cite{1955ApJ...121..161S}.

To ensure accuracy and precision in determining the cluster’s parameters, we adopted search ranges for $\tau$, $[M/H]$, and $dm$ based on prior literature values. Specifically, the $\tau$ was varied between 11 and 13 Gyr with a step size of 0.01 Gyr, the $[M/H]$ ranged from -1.4 to -2.0 dex with a step size of 0.01 dex, and the $dm$ was set to range from 16 to 18 mag with a step size of 0.01 mag. The $[Fe/H]$ was derived from $[M/H]$ using the following formula:
%\begin{linenomath}
\begin{equation}
[Fe/H] = [M/H] - log_{10}(0.694*10^{\alpha/Fe}+0.306),
\end{equation}
%\end{linenomath}
as provided by \cite{2005essp.book.....S}. Here, the value $[\alpha/Fe] = 0.2$ was obtained from spectroscopic analysis performed by \cite{2016A&A...590A...9D}.

Notably, the high stellar density near the core of NGC 5634 led to overexposure and significant dispersion in the faint end of the CMD, as seen in Figure \ref{fig_depth}. To mitigate these effects, we constructed the CMD using photometric data from the region $3\arcmin$ $<$ $r_{cl}$ $<$ $5.73\arcmin$,  avoiding both the overcrowded central core and contamination from field stars in the outer regions. Subsequently, we performed a $\chi^2$ minimization to compare the observed extinction-corrected CMD with theoretical isochrones, where the isochrone parameters were determined for stars with $r$ mag of 14 to 22.5. The model yielding the smallest $\chi^2$ value was identified as the best-fitting isochrone, from which we derived the cluster’s fundamental parameters. The best-fitting parameters for NGC 5634 were determined as follows: $\tau = 12.8 \pm 0.3$ Gyr, $Z = 0.0004 \pm 0.0001$ (corresponding to $[Fe/H] = -1.8 \pm 0.1$ dex and $[M/H] = -1.65 \pm 0.1$ dex, respectively), and $dm = 17.0 \pm 0.1$ mag (equivalent to $d_\odot = 25.12 \pm 1.16$ kpc) (see Figure \ref{fig_cmd1}).

When comparing our results with previous studies, we find good agreement for most parameters. The derived age of $\tau = 12.8 \pm 0.3$ Gyr is consistent with the $\tau = 13$ Gyr estimated by \cite{2021A&A...654A..23G} within the margin of error. However, it differs slightly from the $\tau = 11.84 \pm 0.51$ Gyr reported by \cite{2010MNRAS.404.1203F}, which could be attributed to differences in fitting methods, datasets, and the depth of the observations used in their analysis. The metallicity $[Fe/H] = -1.8 \pm 0.1$ dex is in close agreement with the value of $[Fe/H] = -1.876$ obtained through spectroscopic analysis by \cite{2017A&A...600A.118C}, remaining well within the margin of error. This consistency further validates the robustness of our analysis. For the distance modulus, the derived value of $dm = 17.0 \pm 0.1$ mag aligns well with the mean distance of $dm = 17.07 \pm 0.05$ provided by \cite{2021MNRAS.505.5957B}. It is also comparable to the CMD-derived value of $dm = 17.17 \pm 0.12$ reported by \cite{2002AJ....124..915B}.

\subsubsection{Structural parameters}

The structural parameters of NGC 5634, such as the tidal radius and core radius, have been measured multiple times in the literature \citep[e.g.,][]{1984PASP...96..198K,1989ApJ...339..904C,2005ApJS..161..304M,2012MNRAS.419...14C,2012MNRAS.421..960S} by fitting King models \citep{1962AJ.....67..471K}.  However, some of the reported values appear to exhibit slight discrepancies, which may be attributed to variations in telescope instrumentation and observational conditions. Specifically, \cite{2012MNRAS.421..960S} reported a tidal radius of 5.73$\arcmin$ and a core radius of 0.70$\arcmin$. \cite{1989ApJ...339..904C} estimated a larger tidal radius of 8.35$\arcmin$ and a core radius of 0.21$\arcmin$. In comparison, \cite{2005ApJS..161..304M} provided an even larger tidal radius of 10.37$\arcmin$ and a much smaller core radius of 0.09$\arcmin$.

For our study, we adopted the tidal radius and core radius values of 8.35$\arcmin$ and 0.21$\arcmin$, respectively, as provided by \cite{1989ApJ...339..904C}. This decision was based on our analysis of DESI data, where both CMD-based and proper motion analyses were performed to evaluate the reliability of the three tidal radius estimates. In the CMD analysis, we observed that faint main-sequence stars extended well beyond the smaller tidal radius of 5.73$\arcmin$, albeit with some contamination from field stars. However, we did not detect a significant distribution of main-sequence member stars beyond the tidal radius of 8.35$\arcmin$, supporting the validity of this value. Therefore, to reduce field star contamination and construct a cleaner CMD, we limited the analysis to within a radius of 5.73$\arcmin$ to determine the fundamental parameters. In contrast, the proper motion analysis revealed no statistically significant differences in the proper motion distributions across the three tidal radius intervals, indicating that we cannot conclusively determine the tidal radius based on proper motion analysis alone.

The concentration parameter ($c$) of a GC is defined as the logarithmic ratio of the tidal radius ($r_t$) to the core radius ($r_c$), expressed as:
%\begin{linenomath}
\begin{equation}
c = \log_{10} \left( \frac{r_t}{r_c} \right).
\end{equation}
%\end{linenomath}
Using the tidal radius ($r_t$=8.35\arcmin) and core radius ($r_c$=0.21\arcmin) adopted from \cite{1989ApJ...339..904C}, we calculate the concentration parameter for NGC 5634 as $c$ = 1.60. This value indicates that NGC 5634 has a relatively high concentration. The choice of $r_t$ and $r_c$ is further supported by above CMD analysis, which confirmed that faint main-sequence stars extend up to the tidal radius of 8.35$\arcmin$ but not significantly beyond it, thereby lending confidence to the calculated concentration parameter.

\subsubsection{6D parameters in phase space}

Six-dimensional (6D) phase-space information is essential for calculating the orbit and orbital history of NGC 5634 in various spatial regions, including right ascension (RA), declination (Dec), distance, proper motions in RA and Dec, and radial velocity. These parameters form the foundation for reconstructing the cluster’s orbit and dynamical evolution in the Galaxy.

For the coordinates (RA, Dec), we adopted the values provided by \cite{2019MNRAS.482.5138B}, specifically (217$^{\circ}$.405125, -5$^{\circ}$.976416). The distance to NGC 5634, as determined in this study, is discussed in detail in Section \ref{funpara} and will be used for the subsequent analysis. With these positional parameters established, we will next obtain the 3D velocity components.

Fortunately, Gaia’s high-precision astrometry provides reliable measurements of proper motions and radial velocity. Hence, we adopted the proper motion values reported by \cite{2019MNRAS.482.5138B}, specifically ($\mu_{\alpha}$, $\mu_{\delta}$) = (-1.67, -1.55) mas/yr, with uncertainties of ($\sigma_{\mu_{\alpha}}$, $\sigma_{\mu_{\delta}}$) = (0.02, 0.02) mas/yr. For the radial velocity, we utilized the value of -16.07 km/s, which is also provided by \cite{2019MNRAS.482.5138B}.

\subsection{Matched filter method}\label{MF}

To search for potential extra-tidal structures around NGC 5634, we adopt the matched filtering (MF) method as described by \cite{2002AJ....124..349R}. This method identifies likely cluster members by using the cluster’s CMD as a template. To ensure reliable results and account for 100\% photometric completeness, as discussed in Section \ref{DESIdata}, we use a partial CMD as the MF template. This template is constructed from stars with $g < 23$ and $r < 22.5$ mag, where the completeness remains robust. The analysis is further constrained to stars within the radius range $3.0\arcmin < r_{cl} < 5.73\arcmin$ and with a color index of $-0.5 < g - r < 1.0$. This selection effectively excludes most field stars while retaining the majority of cluster members. Although the tidal radius of the cluster is $8.35\arcmin$, we limit the analysis to $5.73\arcmin$ due to its reduced contamination in this range.

To further minimize contamination from field stars and ensure the CMD template accurately represents the cluster's stellar population, we subtract stars from a normalized reference region located $40\arcmin$ to $50\arcmin$ away from the cluster center. This subtraction removes the contribution of unresolved foreground or background contamination. The resulting Hess difference is then used to define the final cluster CMD template (see Figure \ref{fig_cmd1}). This refined template provides a cleaner and more representative CMD for identifying cluster members.

In addition to the cluster CMD template, we generate a background CMD template for comparison. To construct the background CMD, we select four separate regions, each measuring $1^\circ \times 1^\circ$, located at least $1^\circ$ away from NGC 5634 to avoid overlap with the cluster or its potential tidal structures. These regions are selected to ensure a statistically representative sample of the background stellar population. The number of stars in each background region is normalized to account for differences in area sizes. The background CMD is binned using the same bin size and color-magnitude intervals as the cluster CMD template to ensure consistency in the matched filtering process. These background regions collectively provide a robust CMD template reflecting the realistic distribution of field stars.

After applying the matched filtering method to the data, we extract the density distribution of stars that pass the MF criteria, which we refer to as the $\alpha$ density. The $\alpha$ density represents the matched-filtered signal corresponding to the spatial distribution of stars likely associated with NGC 5634.

\section{Result}
\label{result}

\subsection{Extra-tidal structures around NGC 5634}\label{features}

\begin{figure*}
\centering
\includegraphics[width=0.8\textwidth]{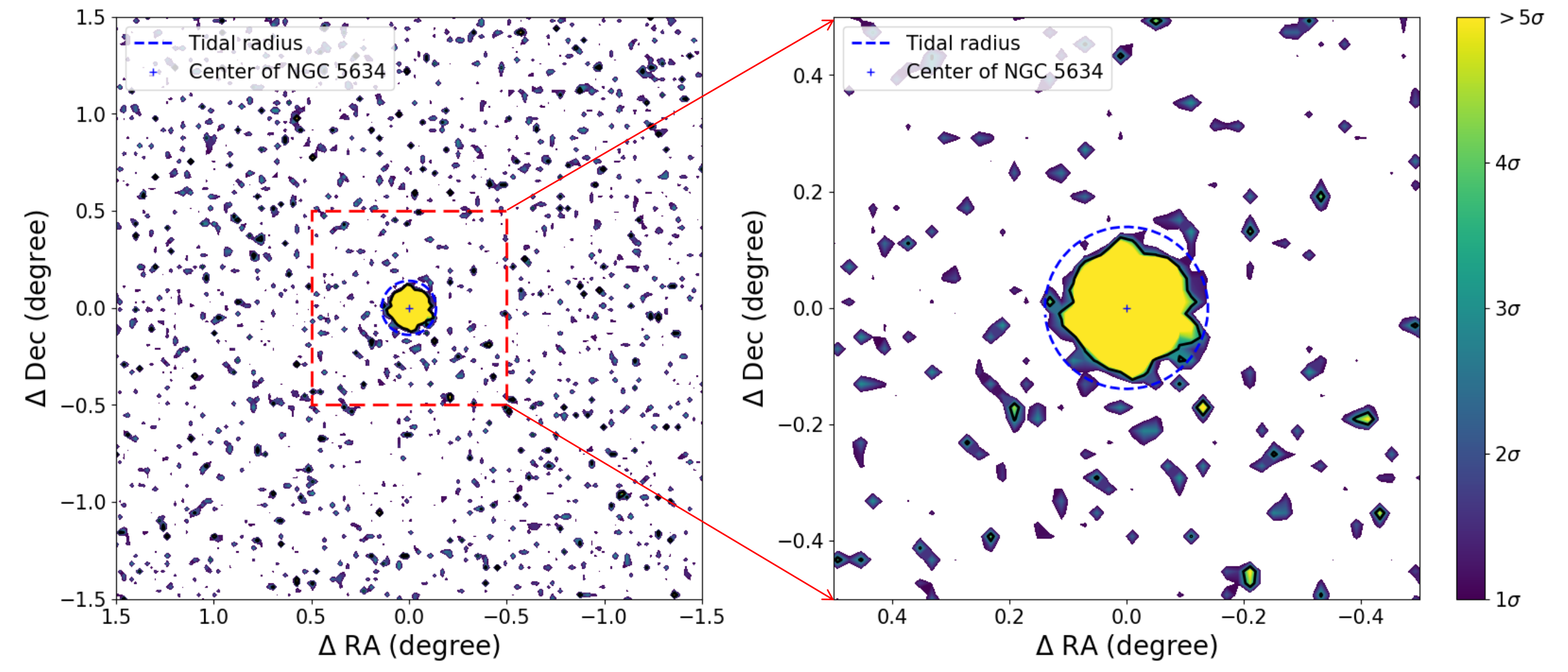}

\caption{SNR distribution centered on NGC 5634 calculated by MF output. The left panel shows the distribution in a 3.0$\times$3.0 square degree space, while the right panel is a close-up view. The blue dashed line in the right panel represents the position of the tidal radius, and the black solid line represents the threshold of 3$\sigma$. Different colors indicate the significance levels as show in the color bar.}
\label{fig_structures}
\end{figure*}

Using the matched filter technique, we obtained the $\alpha$ density distribution centered on NGC 5634, covering an area of $3.0^\circ \times 3.0^\circ$ on the sky. We then applied the following formula to compute the signal-to-noise ratio (SNR) distribution for each bin (0.02$^\circ \times 0.02^\circ$ box on the sky):
%\begin{linenomath}
\begin{equation}
SNR = \frac{\alpha_{{bin}} - \alpha_{{bck}}}{\alpha_{{std}}}.
\end{equation}
%\end{linenomath}
Here, $\alpha_{bin}$ is the value of $\alpha$ for all sources in the bin, $\alpha_{bck}$ is the background value, obtained by averaging after excluding the overly dense regions, 
$\alpha_{std}$ is the standard deviation of the background.  Figure \ref{fig_structures} illustrates the SNR distribution of NGC 5634. The left panel displays the distribution across a $3.0^\circ \times 3.0^\circ$ working field, while the right panel provides a magnified view of a smaller, central region. Our primary goal was to identify potential large-scale extra-tidal structures around NGC 5634. However, despite relaxing the completeness limit of stellar magnitudes, no evidence of extended extra-tidal features, such as tidal tails or extra-tidal stellar overdensities, was detected around NGC 5634. It is likely due to their weak signal blending with the background. This aligns with \cite{2023A&A...673A..44F}, where model fitting revealed the presence of extra-tidal structures with very low intensity, supporting the idea that these faint features may be difficult to distinguish against the background noise.

Based on the study of \cite{2020A&A...637L...2P}, not all galactic GCs exhibit tidal tails or extra-tidal structures, suggesting that their presence possibly depends on specific dynamical and environmental factors. In light of this, we aimed to identify key dynamical parameters that could explain the absence of detectable extra-tidal structures around NGC 5634 and explore its relationship with both the Galaxy and the Sgr dSph.

Firstly, to assess the potential influence of two-body relaxation on NGC 5634, we analyzed the luminosity function (LF) and mass function (MF) of main-sequence stars within the tidal radius of the cluster. We divided the stellar population into three distinct regions: $0\arcmin$ $<$ $r_{cl}$ $<$ $3.0\arcmin$, $3.0\arcmin$ $<$ $r_{cl}$ $<$ $5.73\arcmin$ and $5.73\arcmin$ $<$ $r_{cl}$ $<$ $8.35\arcmin$. Here, the distribution of stars within $3.0\arcmin$ is incomplete, $5.73\arcmin$ corresponds to the smaller tidal radius reported by \cite{2012MNRAS.421..960S}, while $8.35\arcmin$ represents the tidal radius value adopted in this study from \cite{1989ApJ...339..904C}. Subsequently, using the isochrone of NGC 5634 as a reference, we identified main-sequence stars with magnitudes of $r$ $\geq$ 21 mag from the CMD, selecting stars whose color indices fell within three standard deviations ($3\sigma$) of the isochrone. These stars were then used for further analysis. To account for field star contamination, we selected a distant region at $36^\prime \leq r_{cl} \leq 48^\prime$ and subtracted its normalized star counts from those in the three regions. 

\begin{figure*}
\centering
\includegraphics[width=0.4\textwidth]{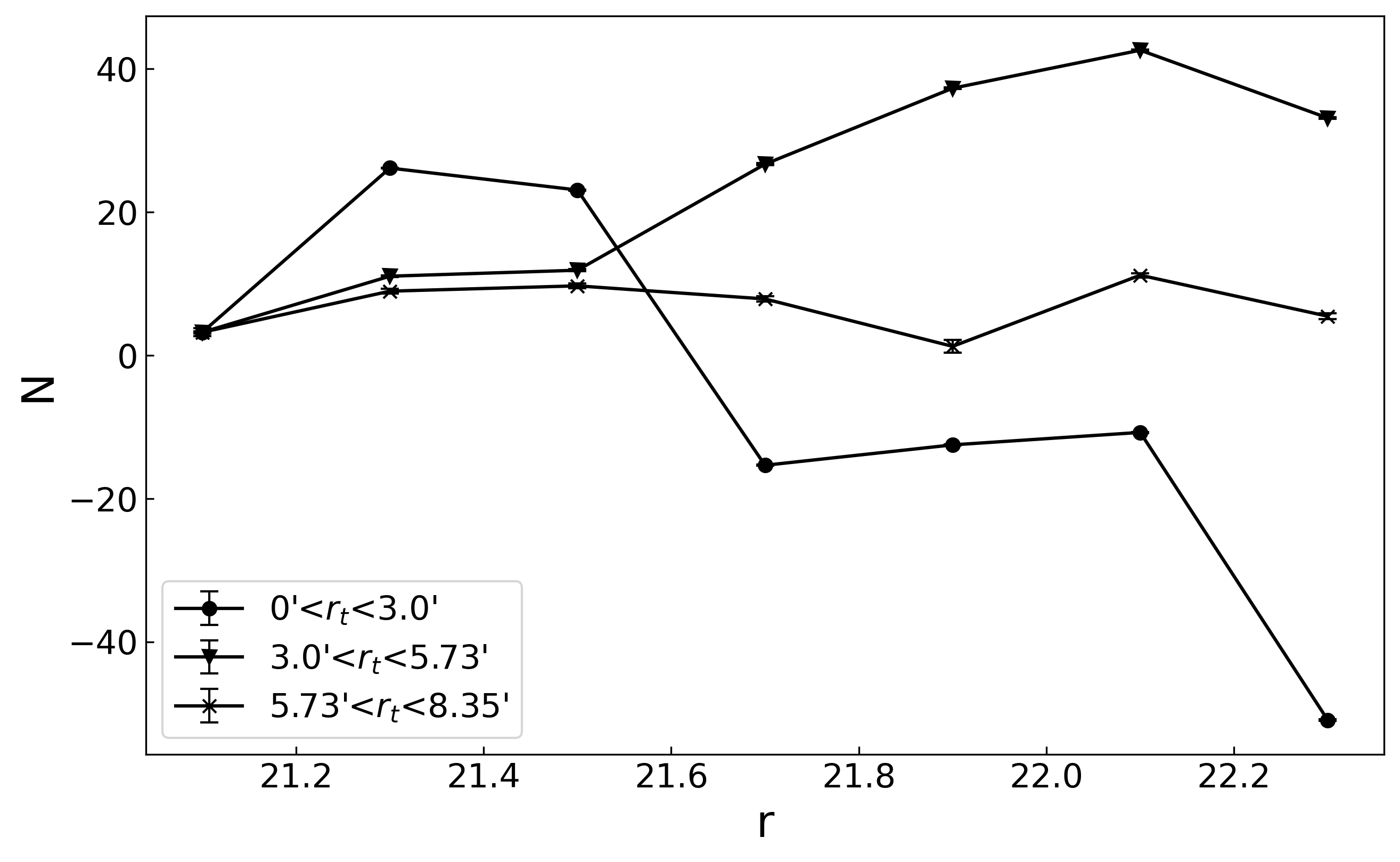}
\includegraphics[width=0.4\textwidth]{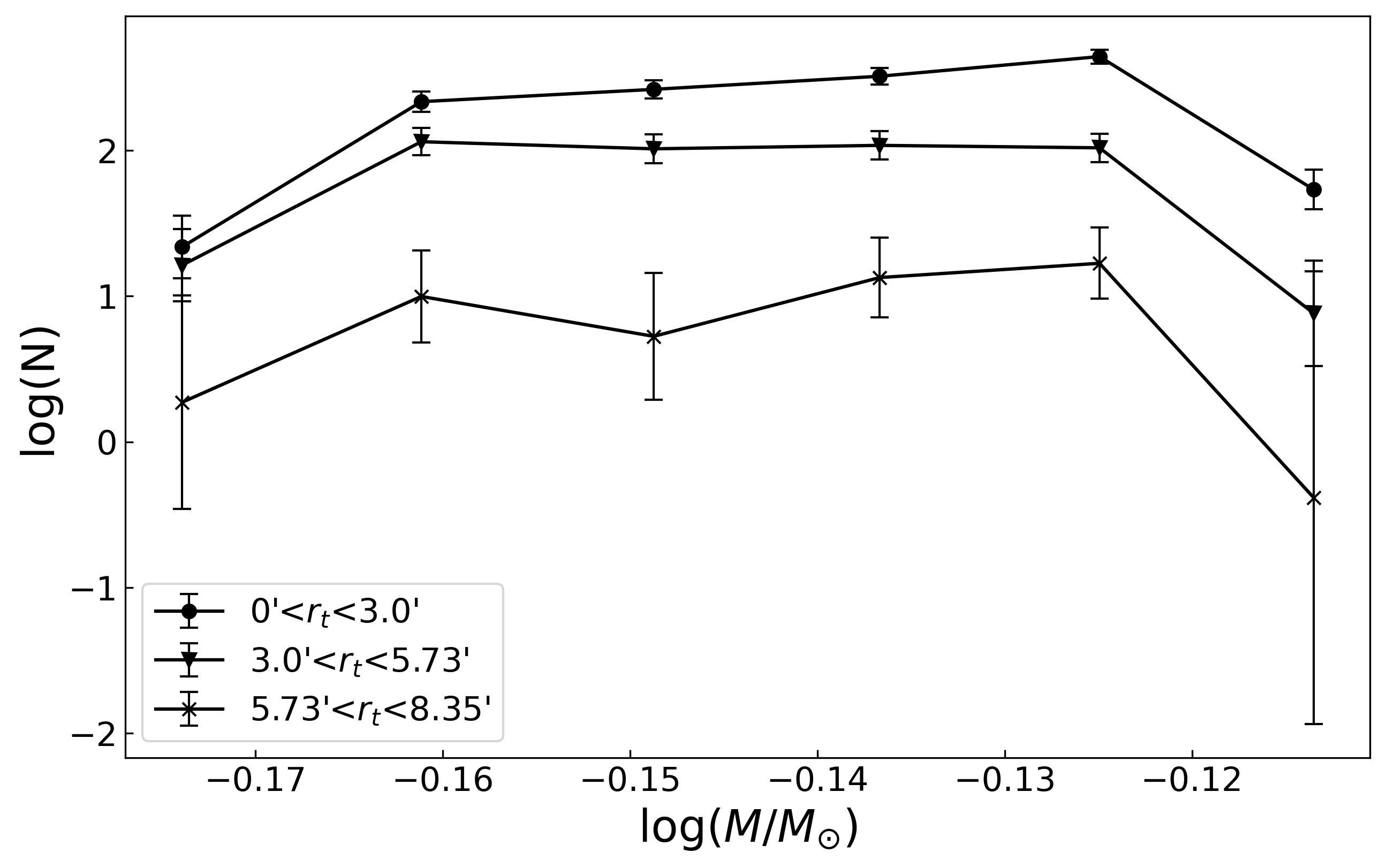}

\caption{Result of mass segregation test. Left panel: LFs in three different regions: $0\arcmin$ $<$ $r_{cl}$ $<$ $3.0\arcmin$, $3.0\arcmin$ $<$ $r_{cl}$ $<$ $5.73\arcmin$ and $5.73\arcmin$ $<$ $r_{cl}$ $<$ $8.35\arcmin$. The star count is normalized at $r$ = 21.1 mag. Right panel: MFs corresponding to the LFs of NGC 5634. The error bars represent 1$\sigma$ Poisson uncertainties derived from the star counts.}
\label{fig_masssegregation}
\end{figure*}

The left panel of Figure \ref{fig_masssegregation} presents the LFs for three different regions, constructed using observed magnitudes. 
For consistency and ease of comparison, the results were normalized to approximately $r = 21.1$ mag. The derived MF was then obtained from the LF by applying the mass-luminosity (M/L) relation based on theoretical isochrone. The absolute magnitudes required for this derivation were obtained by converting the observed magnitudes of the decontaminated stars using a distance modulus of 17.0 mag. The right panel of Figure \ref{fig_masssegregation} presents the MFs for the three different regions. Interestingly, the LF and MF exhibit relatively flat slopes across all regions, with no significant variation. In addition, we conducted a Kolmogorov-Smirnov (K-S) test on the mass distributions of the two completeness regions ($3.0\arcmin$ $<$ $r_{cl}$ $<$ $5.73\arcmin$ and $5.73\arcmin$ $<$ $r_{cl}$ $<$ $8.35\arcmin$). The test yielded a p-value of 0.42, which is greater than the significance level of 0.05, indicating no significant difference between the two distributions. This suggests that mass segregation is not prominent in NGC 5634, which may indicate that the effects of two-body relaxation are relatively weak within the cluster.

Notably, NGC 5634 also exhibits an exceptionally high concentration ($c=1.60$), which is consistent with the value reported by \cite{2002AJ....124..915B}. This suggests that NGC 5634 has a highly compact overall distribution, making it unlikely to possess significant substructure. In addition, the study of \cite{2020A&A...637L...2P} suggests that GCs with high orbital eccentricities (greater than 0.8) and large orbital inclinations (= $\pm$ 70$^\circ$ relative to the galactic plane) are generally more prone to mass loss due to tidal disruption compared to those with more circular orbits and smaller inclinations. From the study of \cite{2019MNRAS.489.4367P}, we obtained the orbital eccentricity of NGC 5634 as e = 0.70 $\pm$ 0.14, the orbital semimajor axis as a = $14.09\pm1.86$ kpc and the orbital inclination as i = 64$^{\circ}$.17 $\pm$ 2$^{\circ}$.87. While these values indicate a moderately eccentric and inclined orbit, they are below the thresholds given by \cite{2020A&A...637L...2P} for enhanced mass loss. This suggests that NGC 5634 is less affected by tidal stripping and has likely retained a substantial fraction of its initial mass. To verify this, we extracted the initial mass ($6.03 \times 10^5 M_{\odot}$) and current mass ($2.47\pm0.48 \times 10^5 M_{\odot}$) of NGC 5634 from the table provided by globular cluster database\footnote{https://people.smp.uq.edu.au/HolgerBaumgardt/globular/}. Subsequently, we calculated the mass loss of NGC 5634 using the formula:
%\begin{linenomath}
\begin{equation}
M_{dis}/M_{ini} = 1/2 - M_{GC}/M_{ini} ,
M_{ini} = M_{GC} + M_{ev} + M_{dis} , 
M_{ev} = 0.5 * M_{ini},
\end{equation}
%\end{linenomath}
as proposed by \cite{2020A&A...637L...2P}, where M$_{ini}$ and M$_{GC}$ represents the initial mass and the current mass respectively, M$_{ev}$ represents the stellar evolution mass, and M$_{dis}$ represents mass lost due to disruption. The exceptionally low mass loss of NGC 5634, estimated at 0.09, suggests that its unique orbital dynamics and kinematic properties may minimize its exposure to the gravitational potential of the Galaxy or its progenitor. The absence of extra-tidal structures limits our ability to assess a morphological link between NGC 5634 and the Sgr stream. Given this, kinematical and dynamical analyses become particularly crucial for understanding its formation and evolution.

%------------------------------------------------
%------------------------------------------------

%------------------------------------------------
%------------------------------------------------

\subsection{Orbital characteristics of NGC 5634, other sources and Sgr stream}
\label{orbit}

\begin{figure*}
\centering
\includegraphics[width=0.8\textwidth]{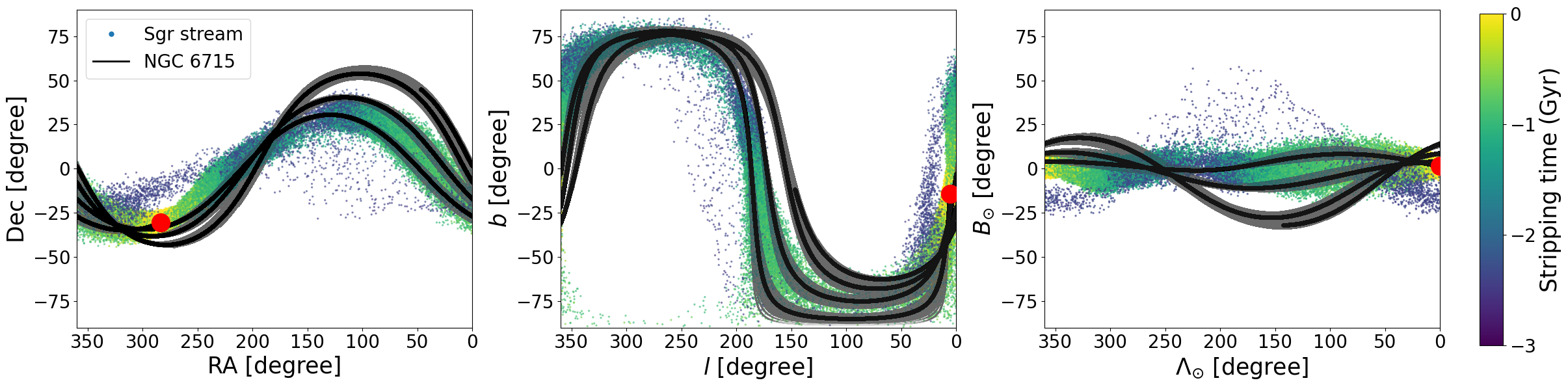}
\includegraphics[width=0.8\textwidth]{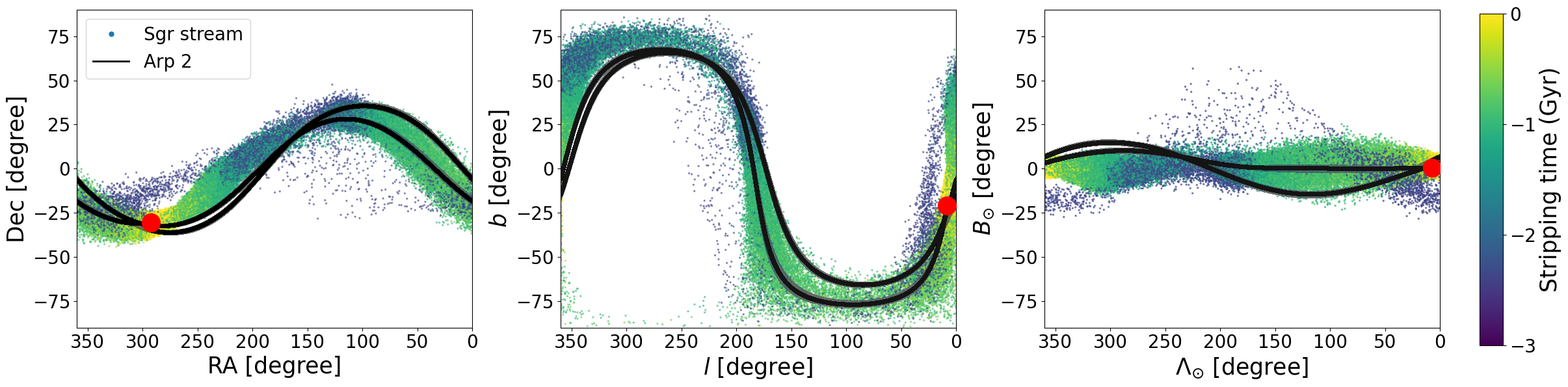}
\includegraphics[width=0.8\textwidth]{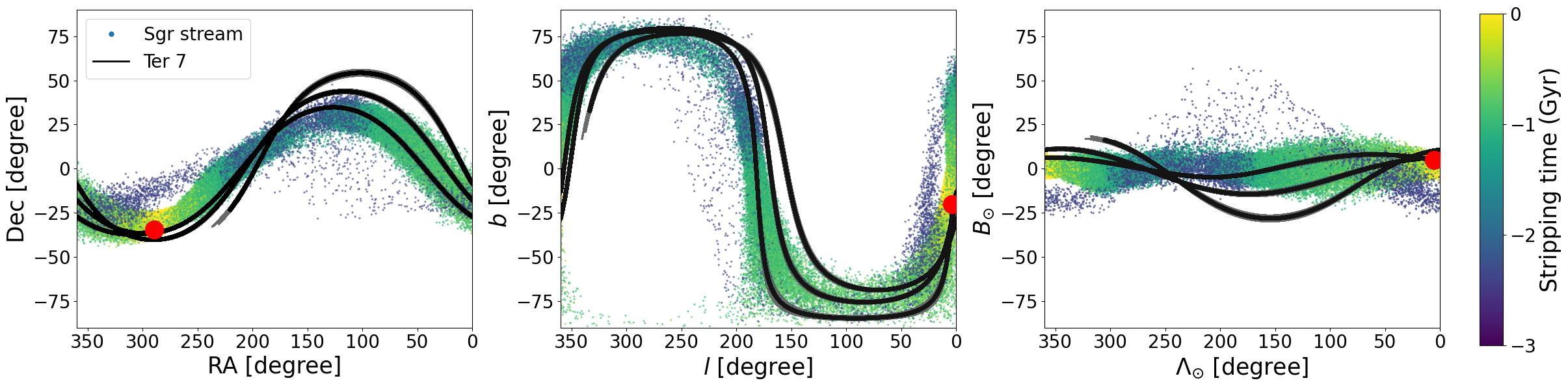}
\includegraphics[width=0.8\textwidth]{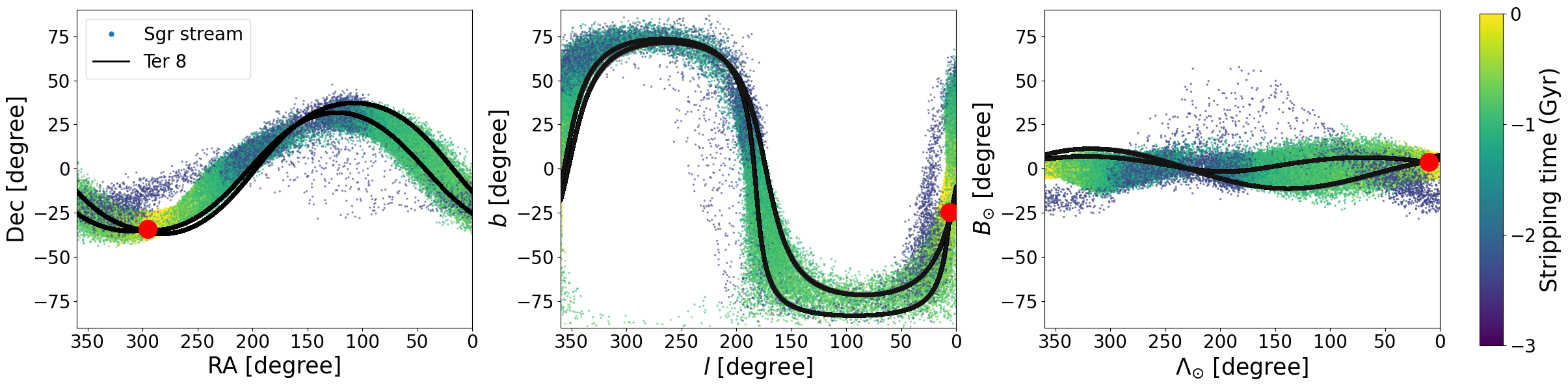}
\includegraphics[width=0.8\textwidth]{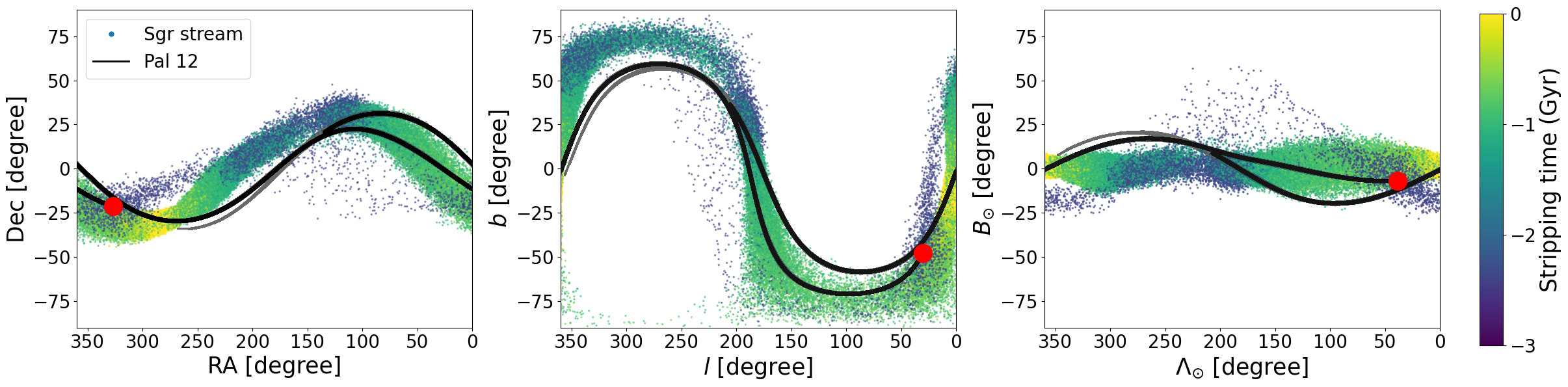}
\includegraphics[width=0.8\textwidth]{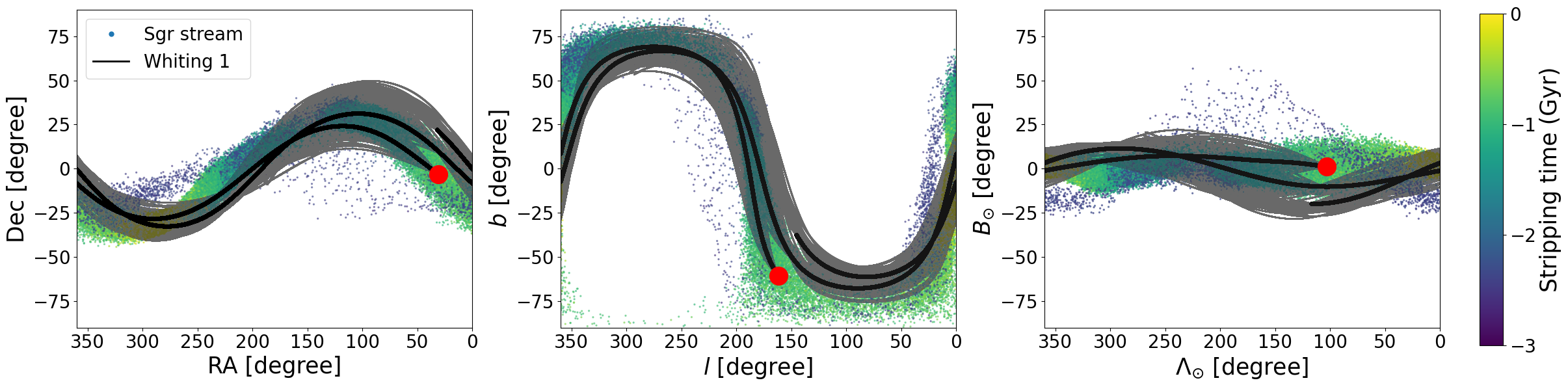}

\caption{Orbit of six confirmed Sgr member clusters with Sgr stream. From left to right panels are the equatorial coordinate system, the galactic coordinate system, and the Sgr coordinate system. The colored scatter points are examples from V21 model, with different colors representing the time at which the particle became unbound from the Sgr dwarf galaxy.} The red solid circle represents the current position of the globular cluster, the black solid line indicates the orbital trace back from the current position, and the gray lines represent the orbits considering the 6D parameter errors.

\label{fig_sixorbit}
\end{figure*}

\begin{figure*}
\centering
\includegraphics[width=0.8\textwidth]{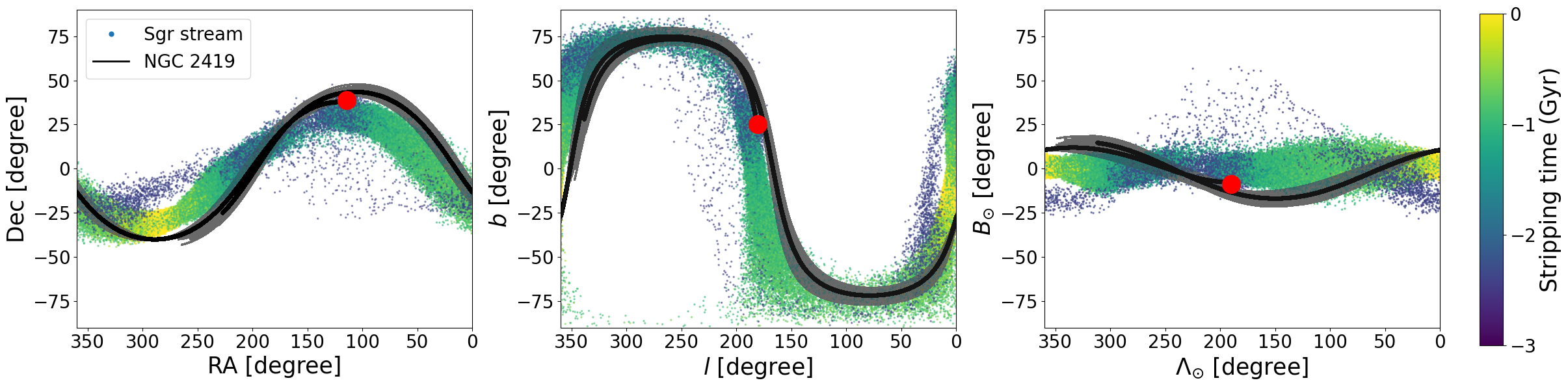}
\includegraphics[width=0.8\textwidth]{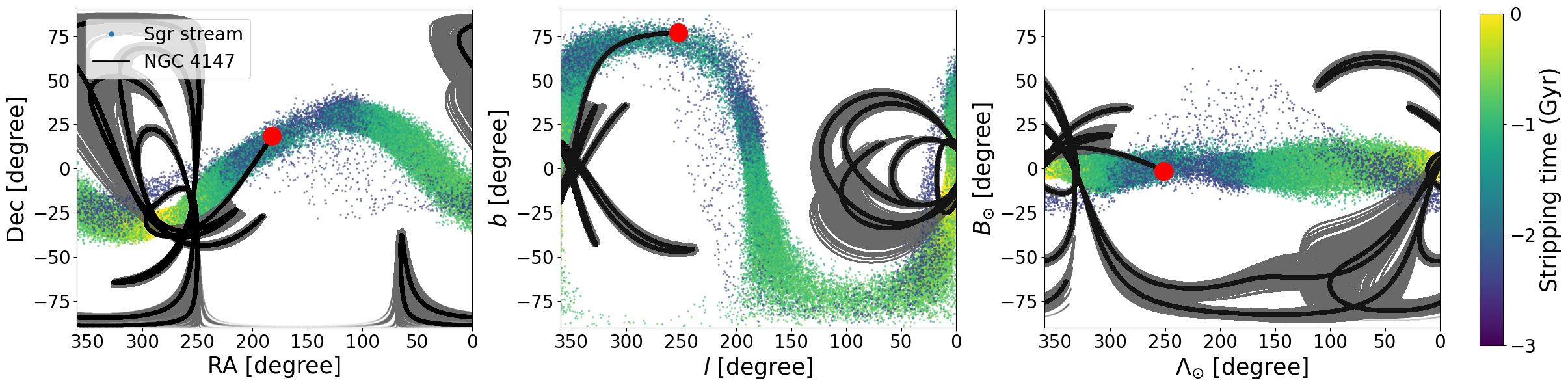}
\includegraphics[width=0.8\textwidth]{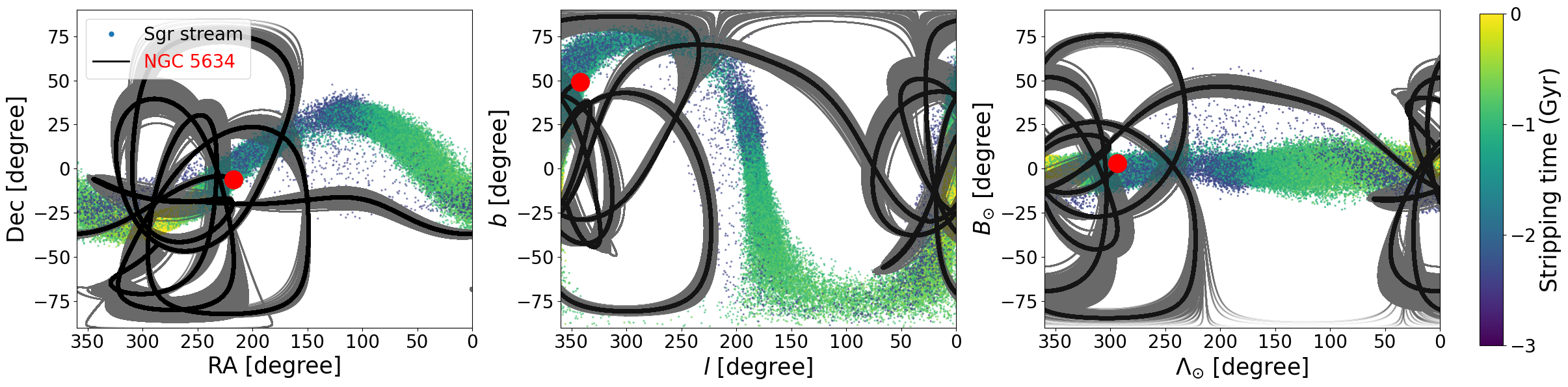}
 
\caption{The same as Figure \ref{fig_sixorbit}, but for orbit of three strong candidate member clusters with Sgr stream. Among them, NGC 5634 is highlighted in red.  }
\label{fig_threeorbit}
\end{figure*}

\begin{figure*}
\centering
\includegraphics[width=0.8\textwidth]{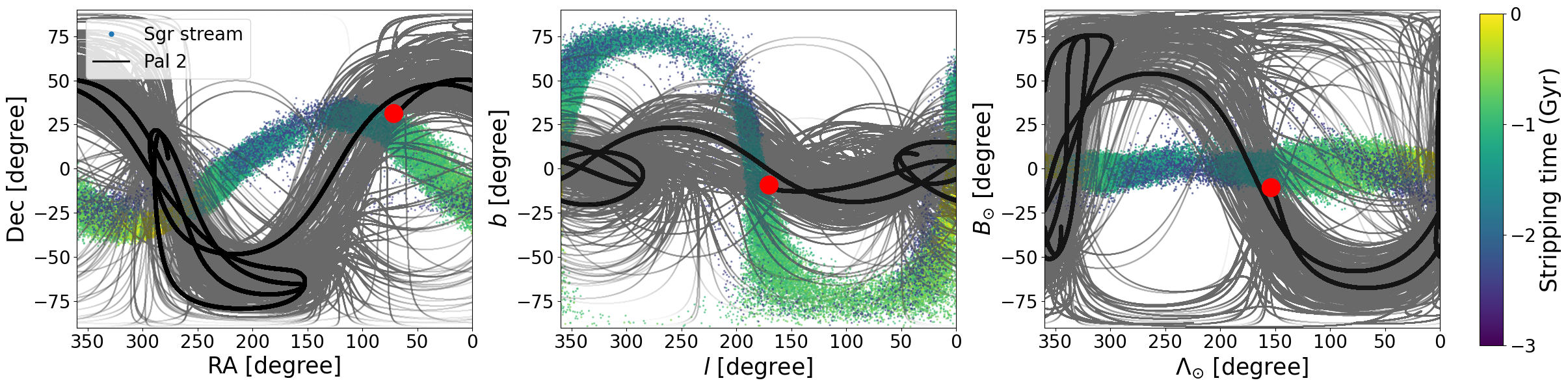}
\includegraphics[width=0.8\textwidth]{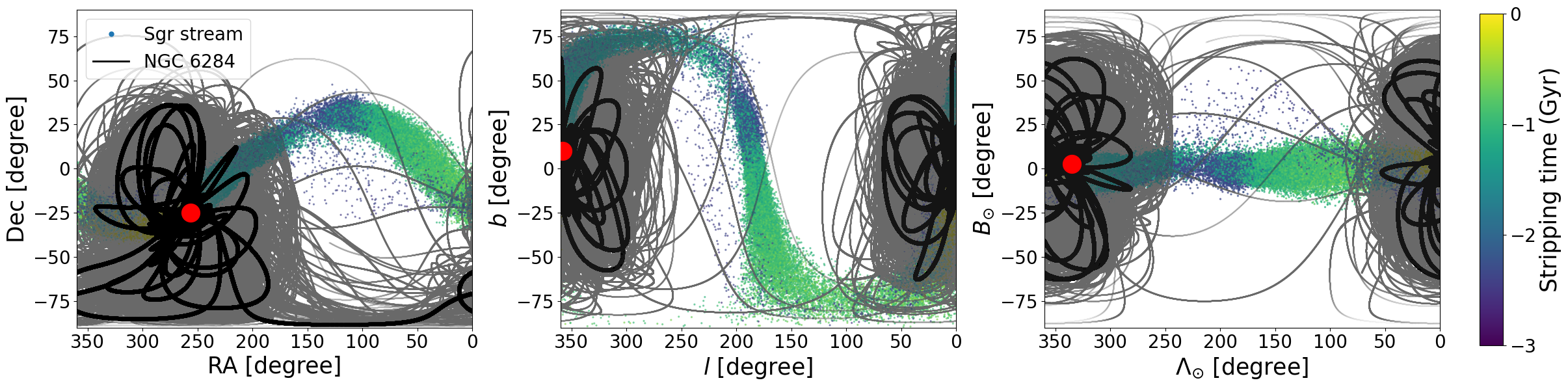}

\caption{The same as Figure \ref{fig_sixorbit}, but for orbit of two weak candidate member clusters with Sgr stream. }
\label{fig_twoorbit}
\end{figure*}

In Section \ref{allpara}, we adopted the 6D phase-space information of NGC 5634 from the literature, enabling detailed calculations of its orbital dynamics and historical orbit in various spatial contexts. Additionally, we extended our study to include orbital calculations for GCs identified as Sgr member GCs, strong candidate GCs, and weak candidate GCs, as provided by \cite{2020A&A...636A.107B}. The 6D phase-space data for these clusters were derived from the dataset provided by \cite{2019MNRAS.482.5138B}. All orbital integrations were carried out with the publicly available AGAMA library\footnote{https://github.com/GalacticDynamics-Oxford/Agama} \citep{2019MNRAS.482.1525V}, which supports multi-component, time-dependent potentials and adaptive Runge-Kutta integration. We employed the evolving triaxial Galactic + LMC potential of \citet{2021MNRAS.501.2279V} (hereafter V21), including both the Large Magellanic Cloud’s perturbation and the halo’s non-axisymmetric shape. Coordinate transformations were performed using the Astropy package \citep{2018AJ....156..123A}, adopting a solar position of $(X,Y,Z)_\odot=(-8.1,0,0.02)$ kpc and a solar motion of $(U,V,W)_\odot=(12.9,245.6,7.8)$ km/s. The orbits of all clusters were integrated backward to $-3$ Gyr to ensure consistency with the Sgr stream model adopted from V21. Moreover, to account for observational uncertainties in cluster velocities and distances, we evaluated potential orbital deviations by generating 100 random values within the reported error margins. The resulting orbital variations are represented by the shaded regions in Figures \ref{fig_sixorbit}, \ref{fig_threeorbit}, and \ref{fig_twoorbit}. 

To investigate the potential association between NGC 5634 and the Sgr stream, we collected spatial distribution data for the stream from the V21 model and compared its orbit with those of the analyzed clusters. Figures \ref{fig_sixorbit}, \ref{fig_threeorbit}, and \ref{fig_twoorbit} illustrate the orbits of Sgr GCs (NGC 6715, Ter 7, Ter 8, Whiting 1, Arp 2, and Pal 12), strong candidate GCs (NGC 2419, NGC 4147, and NGC 5634), and weak candidate GCs (NGC 6284 and Pal 2) in three distinct coordinate systems. Notably, \citet{2024AJ....168..237Z}'s morphological and dynamical studies of NGC 4147 suggest that it may not be associated with the Sgr but instead with the GSE. In this work, we include it in our candidate group for comparative analysis. Among these coordinate systems, the ($\Lambda_{\odot}$, $B_{\odot}$) system is aligned with the orbit of the Sgr stream and is derived by converting the equatorial coordinates ($\alpha$, $\delta$) using the transformations provided by \cite{2014MNRAS.437..116B}:
%\begin{linenomath}
\begin{equation}
\begin{aligned}
\tilde{\Lambda} &= \arctan2(
    -0.93595354 \cos(\alpha) \cos(\delta) 
    - 0.31910658 \sin(\alpha) \cos(\delta) + 0.14886895 \sin(\delta), \\
    &\quad\hphantom{+} 0.21215555 \cos(\alpha) \cos(\delta) 
    - 0.84846291 \sin(\alpha) \cos(\delta) - 0.48487186 \sin(\delta)
)
\end{aligned}
\end{equation}
%\end{linenomath}
%\begin{linenomath}
\begin{equation}
\tilde{B} = \arcsin(0.28103559 \cos(\alpha) \cos(\delta) - 0.42223415 \sin(\alpha) \cos(\delta) + 0.86182209 \sin(\delta)),
\end{equation}
%\end{linenomath}
with $\Lambda_{\odot}=360^{\circ}-\tilde{\Lambda}$, $B_{\odot}=-\tilde{B}$. 

Our results demonstrate that the orbits of the six Sgr GCs closely align with the spatial distribution of the Sgr stream in both the equatorial coordinate system (RA, Dec), galactic coordinate system ($l,b$) and the Sgr-aligned coordinate system ($\Lambda_{\odot}$, $B_{\odot}$). This strong orbital alignment reinforces their classification as members of the Sagittarius. In contrast, among the three strong candidate clusters, only NGC 2419 exhibits a plausible alignment with the stream's distribution. NGC 4147 and NGC 5634 show significant deviations from the Sgr stream in all examined coordinate systems. Similarly, the weak candidate clusters NGC 6284 and Pal 2 also display chaotic and irregular orbital behaviors, akin to those of NGC 4147 and NGC 5634. These deviations may imply that they have different origins, such as the GSE, the Helmi stream, or other accreted systems. Based on this orbital comparison, we find no compelling evidence for a strong association between NGC 5634 and the Sgr system. Therefore, its precise origin remains uncertain and may require further investigation.

%------------------------------------------------
%------------------------------------------------

%------------------------------------------------
%------------------------------------------------

%------------------------------------------------
%------------------------------------------------

The possibility of NGC 5634's association with an ancient tidal stream was first proposed by \cite{2002AJ....124..915B,2003AJ....125..188B}, who highlighted its steep Red Giant Branch and extended blue Horizontal Branch, both strongly resembling those of the Sgr cluster Ter 8. Additionally, the prominent blue Horizontal Branch population closely matches the stellar population observed in the Sgr stream \citep{2002ApJ...569..245N}, and its position and radial velocity are also compatible with that of the Sgr dSph. Subsequent studies focusing on metallicity and chemical abundances have supported its classification as a candidate member of the Sgr stream \citep[e.g.,][]{2010ApJ...718.1128L,2017A&A...600A.118C,2015A&A...579A.104S}. However, alternative investigations using age-metallicity and energy-angular momentum relations suggest that NGC 5634 may instead originate from the Helmi stream, GSE, or other dwarf galaxies, although their focus was not specifically on NGC 5634 \citep[e.g.,][]{2022MNRAS.513.4107C,2023RAA....23a5013S,2022MNRAS.516.4560B,2022ApJ...926..107M,2021MNRAS.508L..26P,2020MNRAS.493..847F}. In fact, the distributions of proper motion, radial velocity, distance, age-metallicity, and [$\alpha/Fe$] for members of the GSE and Helmi streams, including associated clusters, are quite broad, and in some area they overlap. These parameter spaces also overlap partially with that of the Sgr stream. Notably, NGC 5634 is precisely located within these overlapping regions. Therefore, relying solely on velocity, spatial distribution, and age-metallicity relations to determine the cluster's origin may not be entirely reliable.

To distinguish NGC 5634 from GSE or Helmi, we aimed to obtain N-body simulation data for both the GSE and Helmi streams. Although partial data for the Helmi stream were provided by \cite{2019A&A...625A...5K}, they lacked the necessary coverage across multiple coordinate systems, limiting their utility for a comprehensive comparison. Furthermore, we were unable to locate suitable and reliable simulation data for the GSE stream. While we considered using member stars of the GSE and Helmi stream for analysis, their sparse and diffuse distribution proved insufficient for establishing a definitive association with NGC 5634. These limitations highlight the urgent need for more comprehensive data and advanced models to unravel the enigmatic origin of NGC 5634.

\begin{figure*}
\centering
\includegraphics[width=0.6\textwidth]{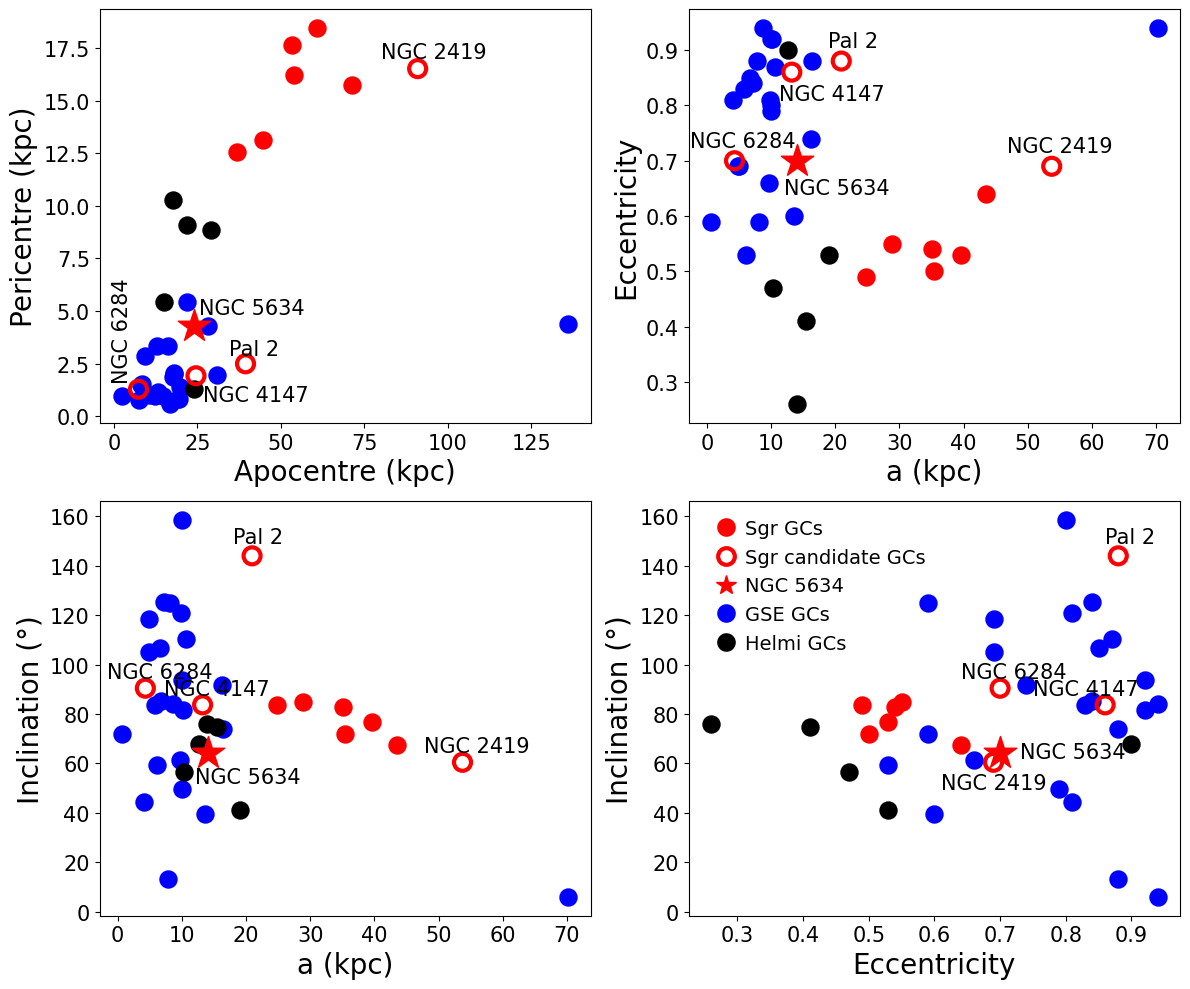}

\caption{Orbital parameter relationships for various GCs. The top-left panel shows the apocentre-pericentre relationship, highlighting the differences in orbital shapes. The top-right panel displays the relationship between orbital eccentricity and semi-major axis (a). The bottom-left panel illustrates the inclination versus semi-major axis, while the bottom-right panel plots orbital inclination against eccentricity. Different groups of GCs are marked with distinct symbols and colors: filled red circles represent confirmed Sgr GCs, open red circles denote strong and weak Sgr candidate GCs, and the red star highlights NGC 5634. Blue dots mark GSE-associated GCs, and black dots represent Helmi stream GCs. Key clusters such as NGC 5634, NGC 4147, NGC 2419, Pal 2, and NGC 6284 are labeled for reference.}
\label{fig_orbitpara}
\end{figure*}

We analyzed globular cluster samples compiled by \cite{2019A&A...630L...4M} and \cite{2020MNRAS.493..847F}, which identify clusters potentially linked to GSE and Helmi progenitor systems. By cross-matching these samples, we created a unified sample that consolidates clusters likely originating from a common progenitor system. We then compared these samples with the 11 clusters analyzed in our orbital study, including NGC 5634. The relationship between the apocentre and pericentre distances of these clusters, based on data provided by \cite{2019MNRAS.482.5138B}, was analyzed in Figure \ref{fig_orbitpara}. The results reveal a distinct clustering pattern among the Sgr GCs, which occupy a specific region in the apocenter-pericenter space. The relatively large pericenter and apocenter distances suggest that these clusters follow wide orbits predominantly residing in the outer Galactic halo, where interactions with the Galactic disk and bulge are limited. This spatial isolation may have contributed to the long-term stability and preservation of their orbital configurations since accretion. Interestingly, NGC 2419, a strong candidate member of the Sgr stream, exhibits a positional similarity to the Sgr clusters. This indicates that its kinematical properties may be closely associated with the Sgr stream. In contrast, clusters associated with the GSE and Helmi streams are primarily concentrated in the lower-left region of the plot, characterized by smaller apocenter and pericenter distances. Notably, NGC 5634 and NGC 4147 are found within this region alongside other clusters, suggesting a potential connection to the GSE or Helmi progenitor system.

In addition, we also examined the relationships between the semi-major axis, orbital eccentricity, and orbital inclination of these clusters using the data provided by \cite{2019MNRAS.489.4367P}, as shown in Figure \ref{fig_orbitpara}. The results also reveal distinct clustering patterns, with different categories of GCs tending to occupy specific regions in this parameter space, suggesting shared kinematical properties among members of the same group. Interestingly, clusters associated with the Sgr stream exhibit relatively longer semi-major axes compared to other clusters, reflecting a wider orbital range. This characteristic is consistent with the perigalacticon and apogalacticon distances. This aligns with the known history of the Sgr system, where its debris trails extend far into the outer halo of the Galaxy, resulting in member clusters having greater galactocentric distances. In contrast, clusters such as NGC 5634 and NGC 4147 display shorter semi-major axes, indicating more confined orbital ranges. This clear kinematical distinction further separates these clusters from the Sgr system and supports the hypothesis that NGC 5634 and NGC 4147 are more likely associated with other galactic substructures, such as the GSE system or the Helmi stream. These clusters also exhibit shorter semi-major axes, which distinguish them from Sgr system.  Moreover, the 6D information, alpha-abundance ratio, and age-metallicity distribution of NGC 5634 overlap significantly with those of the Sgr, GSE, and Helmi systems. Therefore, previous conclusions suggesting that NGC 5634 originated from the Sgr system based solely on these parameters are not sufficiently rigorous.

%------------------------------------------------
%------------------------------------------------

\subsection{Uncertainties Affecting the Cluster's Orbital Path}

The reconstructed orbital path of globular cluster in Section~\ref{orbit} might be subject to several sources of uncertainty, stemming from both model assumptions and observational limitations. In this section, we address factors that may influence the orbital path of a globular cluster and evaluate their impact based on our analysis.
\begin{enumerate}
\item[$\bullet$] Gravitational potential models \\
As described in Section~\ref{orbit}, orbital integrations were carried out using the AGAMA library, which operates based on an input gravitational potential. In order to evaluate the extent to which our orbital integrations are affected by the adopted gravitational potential, we test on three widely used models, i.e., the MWPotential2014 model \citep{2015ApJS..216...29B}, the McMillan2017 model \citep{2017MNRAS.465...76M}, and the V21 model. Our results show that the orbits of the six Sgr clusters remain broadly consistent with the spatial distribution of the Sgr stream under all models. Meanwhile, the remaining four clusters, including NGC 5634, still exhibit significant deviations to the Sgr stream across all three potentials. Therefore, we consider that the differences among these potential models do not affect our conclusions -- namely, the six Sgr clusters maintain agreement with the spatial distribution of the Sgr stream, while the orbits of NGC 5634 and the other three candidate clusters still show significant deviations from the Sgr stream.

\item[$\bullet$] The uncertainty of the 6D phase-data\\
The 6D phase-data are also input parameters of AGAMA. To assess the influence of observational uncertainties in the 6D parameters, we performed orbital integrations incorporating their respective error distributions in Section~\ref{orbit} (see Figures \ref{fig_sixorbit}, \ref{fig_threeorbit}, and \ref{fig_twoorbit}). The resulting orbits for the six Sgr clusters still exhibit well aligned with the spatial distribution of the Sgr stream. A comparable level of consistency is also observed for the strong candidate NGC 2419. In contrast, the remaining four candidate clusters, including NGC 5634, do not show a good match with the Sgr stream within the uncertainty range. Therefore, the uncertainties arising from observational errors appear to have limited impact on the orbital results.

However, readers should be aware that if a substantially different set of values - or even individual parameters - is used for the 6D information such as distance, proper motion, or radial velocity, the resulting orbital trajectories may change accordingly, and this applies to all clusters. Nevertheless, the 6D information we adopted is taken from well-established and high-precision studies, and the values are consistent with observational results reported in various independent sources. Therefore, we consider the orbital trajectories presented in this work to be reliable.

\item[$\bullet$] The mass evolution of the progenitor\\
It should also be emphasized that the currently observed phase-space information of the GCs may not fully represent the dynamical properties of the stream stars, as this discrepancy could be influenced by the mass evolution of the progenitor. If a GC was stripped at an early stage, it would have been affected by the gravitational potential of a more massive progenitor than the remnant observed today. As a result, the present-day kinematics of NGC 5634 could differ somewhat from those of the stellar component of the stream, and thus a direct comparison should be interpreted with some caution.

A quantitative assessment of the impact of this effect would require highly complex modeling of the progenitor’s mass evolution and tidal stripping processes, which is beyond the scope and capacity of this study. The comparisons presented here are based on the best currently available models and kinematic data. While we acknowledge that uncertainties in the progenitor’s mass - and its potential variation over time - may influence the results to some extent, we consider our orbital comparisons to be informative under relatively idealized assumptions.
\end{enumerate}

\subsection{Total orbital energy and angular momentum distribution of NGC 5634 and other sources}

\begin{figure}
\centering
\includegraphics[width=0.5\textwidth]{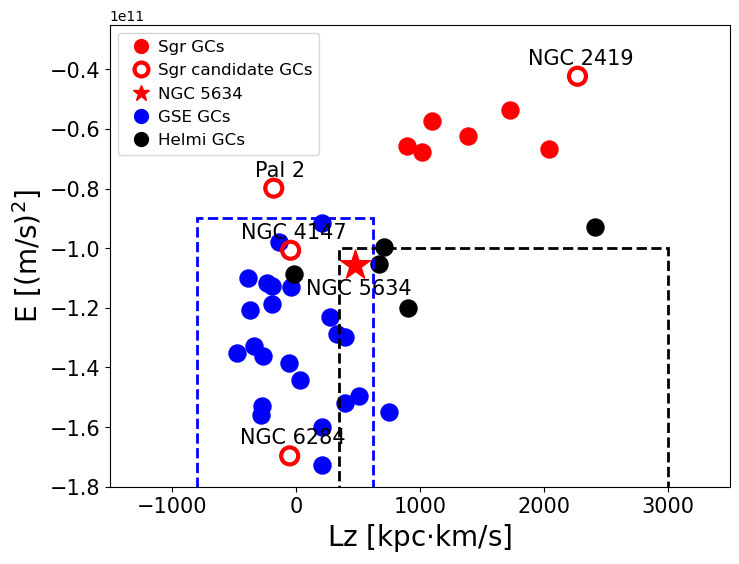}

\caption{
Relationships between the angular momentum along z direction Lz vs. total orbital energy E. Different symbols and colors correspond to various cluster populations based on their potential origins: confirmed Sgr clusters are marked with filled red circles, strong and weak Sgr candidate clusters are denoted by open red circles, and the red star highlights NGC 5634. Clusters associated with the GSE system are represented by blue dots, while those associated with the Helmi stream are represented by black dots. The blue and black dashed boxes represent the GSE and Helmi regions, respectively. Key clusters are labeled for reference, the same as in Figure \ref{fig_orbitpara}.
}
\label{fig_EL}
\end{figure}

To further investigate the potential origin and evolutionary history of NGC 5634, we analyzed its orbital angular momentum along the z-axis (Lz) and total orbital energy (E) in comparison with other sources, as shown in Figure \ref{fig_EL}. The calculation of energy and angular momentum is based on the same 6D information provided by \cite{2019MNRAS.482.5138B}, while the parameter space of GSE and Helmi are derived from \cite{2019A&A...630L...4M}. This dynamical approach allows us to further evaluate NGC 5634’s potential association with the Sgr stream, the Helmi stream, or GSE merger remnants.

Our results indicate that Sgr clusters exhibit a distinct clustering in the Lz-E plane, reflecting their shared origin within the gravitational potential of the Sgr dwarf galaxy. These clusters possess relatively high orbital energy and angular momentum, consistent with the known dynamical history of the Sgr system. Similarly, NGC 2419, as a strong candidate member of the Sgr stream, is located near this cluster group. Its Lz and E values closely align with the characteristics of confirmed Sgr members, further supporting its classification as a potential member of the Sgr stream. The positions of NGC 5634 and NGC 4147 are distinctly separated from the main Sgr group. Both clusters exhibit intermediate Lz and E values, indicating that their dynamical origins are fundamentally different from the Sgr system. NGC 5634 lies within both the blue and black dashed regions, overlapping with the GSE and Helmi stream parameter spaces defined by \cite{2019A&A...630L...4M}. This ambiguity brings uncertainty to the origin of NGC 5634. On the other hand, NGC 4147 is located squarely within the GSE-defined region, suggesting a strong likelihood of belonging to the GSE system. The lack of clustering with the Sgr member group further supports the conclusion that both clusters are dynamically distinct from the Sgr stream. The positions of NGC 6284 and Pal 2 in the Lz-E plane also differ significantly from the confirmed Sgr clusters. NGC 6284 is located closer to the GSE-defined region, suggesting that its orbital properties are more consistent with the GSE system rather than the Sgr stream. On the other hand, Pal 2 exhibits relatively high orbital energy but does not clearly align with any specific group, indicating a potentially complex dynamical history.

%------------------------------------------------
%------------------------------------------------

\section{Summary}\label{summary}

We investigated the properties and origin of NGC 5634 primarily from three perspectives: morphology, kinematics, and dynamics. First, using the deep data from the DESI Legacy Survey, we redetermined the fundamental parameters of NGC 5634 (age, metallicity, and distance) by fitting theoretical isochrones. Additionally, we utilized the deep survey data and applied the matched-filter method to detect extra-tidal structures surrounding NGC 5634. Furthermore, we reconstructed its orbital history using 6D phase-space data to trace its orbit through the galactic potential, and discussed some uncertainties associated with this process. Finally, we analyzed its orbital parameters, including its semi-major axis, eccentricity, inclination, and apocenter-pericenter distances, to characterize its orbital properties. By mapping the clusters onto the Lz–E plane, we further analyzed their population distribution characteristics and potential origins. Based on this comprehensive analysis, we arrived at the following results:

1. No significant structural features beyond the tidal radius of NGC 5634 were detected above the $3\sigma$ level, even without accounting for data completeness. Within the tidal radius, the absence of mass segregation suggests minimal two-body relaxation, likely suppressing the formation of extra-tidal features. Overall, the presence or absence of such structures appears to result from a complex interplay of factors—including mass segregation, concentration, orbital inclination, and eccentricity—rather than any single dominant mechanism.

2. Through a comparison with the orbits of Sgr member clusters, we found that the orbits of Sgr clusters remain consistent with the spatial distribution of the Sgr stream over long-term integrations. In contrast, the orbit of NGC 5634 only briefly intersects with the Sgr stream and shows significant divergence when traced further back in time. Based on this result, we suggest that there is currently insufficient evidence to support a strong association between NGC 5634 and the Sgr system. However, we also discussed several factors that could potentially affect the orbital evolution, as well as the potential limitations and risks of relying solely on direct orbit comparisons.

3. The Sgr GCs tend to have more distant pericentre and apocentre distributions, often accompanied by extended orbital ranges and relatively stable orbits. In contrast, NGC 5634 shares greater similarities with GSE or Helmi-related clusters, which are characterized by closer pericentre and apocentre distributions and more compact, dynamically evolving orbits.

4. The position of NGC 5634 in the Lz–E plane lies closer to the parameter space occupied by GSE and Helmi clusters. Its total orbital energy and angular momentum differ significantly from those of confirmed Sgr clusters. This distinct dynamical location suggests that NGC 5634 is more likely associated with the GSE, the Helmi stream, or another progenitor system.

These findings contribute to the discussion of NGC 5634’s origin and provide new insights into assessing the potential
association of GCs with the Sgr stream based on their kinematical and dynamical properties. Based on these results, we suggest
that NGC 5634 is unlikely to have originated from the Sgr system and is more likely associated with the GSE or Helmi
stream. Future studies aiming to accurately determine the origins of such clusters, particularly those lacking extra-tidal
structures, will require higher-precision data, better evolution models and the development of new analytical methods.

\acknowledgments
Jundan Nie acknowledges the support of the National Key R\&D Program of China (grant Nos. 2021YFA1600401 and 2021YFA1600400), the Chinese National Natural Science Foundation (grant No. 12373019), the Beijing Natural Science Foundation (grant No. 1232032), and the science research grants from the China Manned Space Project (grant Nos. CMS-CSST-2021-B03, CMS-CSST-2021-A10 and CMS-CSST-2025-A11). Biwei Jiang acknowledges the support of the Chinese National Natural Science Foundation (grant No. 12133002). Hao Tian acknowledges the support of the National Key R\&D Program of 
China (No. 2024YFA1611902).

\newpage
\bibliography{wangszGC}
\bibliographystyle{aasjournal}

\end{document}